\documentclass[twocolumn,pra,aps,showpacs,floatfix]{revtex4}
\usepackage{epsfig}

\def\beq{\begin{equation}}
\def\eeq{\end{equation}}
\def\bea{\begin{eqnarray}}
\def\eea{\end{eqnarray}}

\newcommand{\la}[1]{\label{#1}}
\newcommand{\ur}[1]{(\ref{#1})}

\newcommand{\eq}[1]{eq.~(\ref{#1})}
\newcommand{\eqs}[2]{eqs.~(\ref{#1},\ref{#2})}
\newcommand{\eqss}[2]{eqs.~(\ref{#1}-\ref{#2})}
\newcommand{\Eq}[1]{Eq.~(\ref{#1})}

\newcommand{\Eqss}[2]{Eqs.~(\ref{#1}-\ref{#2})}
\newcommand{\half}{\frac{1}{2}}

\newcommand{\at}{\overline{10}}

\newcommand{\dec}{\left({\bf 10},\frac{3}{2}\right)}

\newcommand{\n}{\nonumber}
\begin{document}
\title{Where are the missing members of the baryon
antidecuplet?}
\author{Dmitri Diakonov$^{* \diamond}$ and
Victor Petrov$^*$}

\affiliation
{$^*$ St. Petersburg Nuclear Physics Institute,
Gatchina, St. Petersburg 188300, Russia \\
$^\diamond$ NORDITA, Blegdamsvej 17, DK-2100 Copenhagen \O,
Denmark}

\date{October 17, 2003}

\begin{abstract}
We analyze what consequences has the observation of exotic
pentaquark baryons on the location of the non-exotic baryons
belonging to the antidecuplet. We suggest that there must be a new
nucleon state at 1650-1690 MeV and a new $\Sigma$ baryon at
1760-1810 MeV.
\end{abstract}


\maketitle

\noindent
The last 12 months have witnessed a dramatic development in baryon
spectroscopy. First, an ``exotic" baryon $\Theta^+$ with strangeness
$+1$ has been discovered \cite{Osaka,ITEP,JLab,ELSA,Asratyan} whose
mass and narrow width is in agreement with the theoretical prediction
\cite{97}. Quite recently, the first observation of a double-strange
quadruplet of baryons $\Xi_{\frac{3}{2}}$ has been reported
\cite{CERN}. Both $\Theta^+$ and $\Xi_{\frac{3}{2}}$ have quantum
numbers such that they can be `made of' minimally four quarks and an
antiquark, hence dubbed pentaquarks. Both $\Theta^+$ and
$\Xi_{\frac{3}{2}}$ are expected in the flavor $SU(3)$ antidecuplet of
baryons with spin and parity $\half^+$\cite{97}, forming the upper vertex
and the lower edge of the
big triangle in the isospin-strangeness diagram, see Fig.~1.

The antidecuplet implies that there are also non-exotic baryons in the
middle of the triangle, with quantum numbers of a nucleon and a
$\Sigma$ hyperon. These particles have not been identified yet. Finding
them is important for understanding the dynamics leading
to pentaquarks. In this note, we examine the overall panorama of
baryons below 2 GeV with spin and parity $\half^+$ and suggest
where the antidecuplet $N$ and $\Sigma$ can be expected.

To simplify the discussion we put $m_u=m_d=0$ and neglect isotopic
splittings throughout the paper. If in addition one takes $m_s=0$ all
hadrons appear in exactly degenerate multiplets of the $SU(3)$ flavor
group. The splittings inside multiplets and mixings of particles from
different multiplets are due to $m_s\neq 0$. In QCD, the only source of
$SU(3)$ violation is the mass term of the Lagrangian,
$m_s\bar ss$, which can be written as
$\bar q\left(\frac{m_s}{3}
{\bf 1}_3 -\frac{m_3}{\sqrt{3}}\lambda_8\right)q$. The first
(singlet) part leads neither to splitting nor mixing.
The second part transforms as the
$8^{\rm th}$ component of the octet. To get the splitting/mixing mass
matrix, one sandwiches it between the baryon states in question,
$m_s\!<\!\!B|\bar q\lambda_8 q|B'\!\!>$. If $B,B'$ belong to
octets this matrix element is characterized by two constants, as there
are two ways to extract an octet from a direct product of $8\otimes 8$.
If $B,B'$ are members of the decuplet or antidecuplet, there
is only one constant, as there is a unique way to get the decuplet
(antidecuplet) from the product of $10(\at)\otimes 8$. If the mixing
between an octet and an antidecuplet is considered there is also only
one constant, as there is only one way to get an antidecuplet from
$8\otimes 8$.

If one temporarily disregards mixings of multiplets, the mass matrices
for an octet and a decuplet are:
\beq
\begin{array}{ccc}
N_8&=&M_8-\frac{7}{4}x-y,\\
\Lambda_8&=&M_8-x,\\
\Sigma_8&=&M_8+x,\\
\Xi_8&=&M_8+\frac{3}{4}x+y,
\end{array}\qquad
\begin{array}{ccc}
\Delta_{10}&=&M_{10}-y^\prime,\\
\Sigma_{10}&=&M_{10},\\
\Xi_{10}&=&M_{10}+y^\prime,\\
\Omega_{10}&=&M_{10}+2y^\prime.
\end{array}
\la{split810}\eeq
The masses of antidecuplet baryons are also equidistant,
\beq
\begin{array}{ccc}
\Theta^+&=&M_{\at}-2z,\\
N_{\at}&=&M_{\at}-z,\\
\Sigma_{\at}&=&M_{\at},\\
\Xi_{\frac{3}{2}}&=&M_{\at}+z.
\end{array}
\la{split_at}\eeq
The exotic members of the antidecuplet cannot mix up with octet
baryons; in principle they can mix up with even higher exotic
multiplets but we ignore this possibility. Therefore, the
center $M_{\at}$ and the spacing $z$ are found from $\Theta$ and
$\Xi_{\frac{3}{2}}$ masses. We use $\Theta=1539\pm2\,{\rm MeV}$
\cite{ITEP} and $\Xi_{\frac{3}{2}}=1862\pm2\,{\rm MeV}$ \cite{CERN}
and find $M_{\at}=(\Theta+2\Xi)/3\approx 1754\,{\rm MeV}$ and
$z=(\Xi-\Theta)/3\approx 108\,{\rm MeV}$. Therefore, if $N_{\at}$
and $\Sigma_{\at}$ do not mix up with other states their masses are
\beq
N_{\at}\approx 1647\,{\rm MeV},\qquad
\Sigma_{\at}\approx 1755\,{\rm MeV}.
\la{NSzero}\eeq

A consequence of \eq{split810} is the Gell-Mann--Okubo relation for
octet masses:
\beq
\frac{N_8+\Xi_8}{2}=\frac{3\Lambda_8+\Sigma_8}{4}.
\la{GMO1}\eeq
For the ground-state octet, the least-square fit of the masses
averaged over isospin components gives $M_8=1151.5\,{\rm MeV},\,
x=40.2\,{\rm MeV},\, y=139.3\,{\rm MeV},\,\sqrt{\sigma^2}=3.08\,{\rm
MeV}$ with GMO relation satisfied with 0.5\% accuracy.  The equal
spacing in the decuplet $\dec$ is also satisfied with high accuracy,
with the best fit to the masses $M_{10}=1382.1\,{\rm
MeV},\,y^\prime=147.0\,{\rm MeV},\,\sqrt{\sigma^2}=3.45\,{\rm
MeV}$. The success of this 40-year-old exercise is an unambiguous indication
that the mixings of the lowest octet and decuplet with other
multiplets are small, as are other possible higher-order $m_s$
corrections to the masses. The GMO relation is valid only in the
linear order in $m_s$ and thus it serves as a test for higher-order
corrections. Let us apply this test to higher $\half^+$ multiplets.

The Particle Data Group suggests candidates
filling precisely two $\half^+$
octets in the mass range $1\!-\!2\,{\rm GeV}$. One is
$N(1710),\,\Lambda(1810),\,\Sigma(1880)$ and $\Xi(1950)$. The spin
and parity of the last baryon are unknown but
we assume that they
are $\half^+$. These masses satisfy the GMO relation \ur{GMO1} with
an astonishing accuracy of $0.3\%$. It can be accidental, given
that the masses are not known with that precision, however it
indicates that it is anyhow a rather ``pure'' octet with not much
room left for its mixing with other multiplets. The situation with
the second octet formed by $N(1440),\,\Lambda(1600),\,\Sigma(1660)$
and $\Xi(1690)$ is worse: taken at face value these numbers satisfy the
GMO relation \ur{GMO1} only with an accuracy of 3\%. We stress that
no other $\half^+$ particles are mentioned by the PDG; those that
appear in the baryon listings fall apparently into these two octets.
One octet does not seem to allow considerable mixing; the other can
in principle mix with other multiplets, in particular with the
antidecuplet.

Before presenting the general scheme of antidecuplet-octet mixing,
let us recall the standard theory of the octet-singlet mixing of
mesons (see {\it e.g.} \cite{Kokk}). Denoting $\theta$ the
mixing angle, the GMO relation \ur{GMO1} is modified:
\beq
m_K^2=\frac{3(m_1^2\cos^2\!\theta+m_2^2\sin^2\!\theta)
+m_{I\!=\!1}^2}{4},
\la{GMOmes}\eeq
where $m_1(m_2)$ are the masses of the lighter (heavier)
iso\-singlet mesons, $m_K$ is the mass of the strange and
$m_{I\!=\!1}$ is the mass of isotriplet members of the octet.
\Eq{GMOmes} works well for pseudoscalar, vector and tensor
mesons, with the mixing angles $\theta\!=\!-\!\!10.6^{\rm\small o},
40^{\rm\small o}, 30.2^{\rm\small o}$, respectively.

\begin{figure}
\begin{picture}(100,150)
\put(-80,0){
\epsfxsize=8.5cm
\epsfysize=5.5cm
\epsfbox{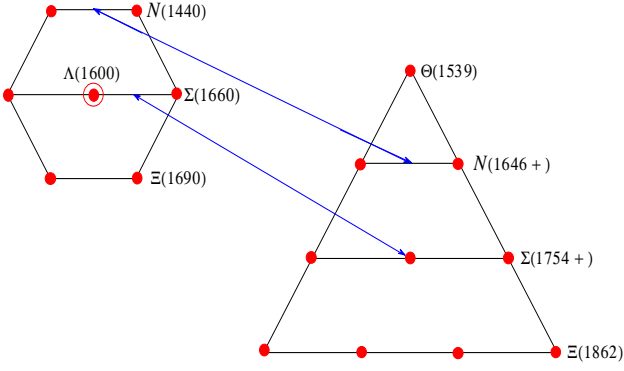}}
\end{picture}
\caption{Possible octet-antidecuplet mixing.}
\end{figure}

Let us now derive the general equations for octet-antidecuplet
mixing, analogous to \eq{GMOmes}. There are two particles
common to the octet and the antidecuplet: $N$ and $\Sigma$, and only
they can mix up. Let us denote by $N_8,N_{\at}$ and
$\Sigma_8,\Sigma_{\at}$ the would-be masses of the appropriate
particles before the mixing. These masses include the
corrections linear in $m_s$, as given by \eqs{split810}{split_at}.
In particular, $N_{\at}=1646\,{\rm MeV},\;\Sigma_{\at}=1754\,{\rm MeV}$
according to \eq{NSzero}; $N_8$ and $\Sigma_8$ are {\it a priori}
unknown.  Let the mixing matrix element be
$V=<\!N_{\at}|m_s\bar s s|N_8\!>=$
$<\!\Sigma_{\at}|m_s\bar s s|\Sigma_8\!>$.
It is important that the two matrix elements are equal
since the two corresponding $SU(3)$ Clebsch--Gordan coefficients are
equal \cite{deSwart}. The physical states are obtained from the
diagonalization of the two $2\times 2$ matrices:
\beq
\left(
\begin{array}{cc} N_8&V\\V&N_{\at} \end{array}\right), \qquad
\left(
\begin{array}{cc} \Sigma_8&V\\V&\Sigma_{\at}
\end{array}\right),
\la{mmatrices}\eeq
with the result
\beq
\begin{array}{l}
|N_1\rangle=|N_8\rangle\cos\theta_N+|N_{\at}\rangle\sin\theta_N,\\
|N_2\rangle=-|N_8\rangle\sin\theta_N+|N_{\at}\rangle\cos\theta_N,
\end{array}\quad \tan2\theta_N\!=\!\frac{2V}{N_8-N_{\at}},
\la{Nphys}\eeq
\beq
\begin{array}{l}
|\Sigma_1\rangle=|\Sigma_8\rangle\cos\theta_\Sigma
+|\Sigma_{\at}\rangle\sin\theta_\Sigma,\\
|\Sigma_2\rangle=-|\Sigma_8\rangle\sin\theta_\Sigma
+|\Sigma_{\at}\rangle\cos\theta_\Sigma,
\end{array}
\quad
\tan2\theta_\Sigma\!=\!\frac{2V}{\Sigma_8-\Sigma_{\at}},
\la{Sphys}\eeq
and physical masses
\bea
\la{N12}
N_{1,2}&=&\frac{1}{2}\left(N_{\at}+N_8\mp
\sqrt{(N_{\at}-N_8)^2+4V^2}\right),\\
\la{S12}
\Sigma_{1,2}&=&\frac{1}{2}\left(\Sigma_{\at}+\Sigma_8\mp
\sqrt{(\Sigma_{\at}-\Sigma_8)^2+4V^2}\right).
\eea
The ``bare'' octet and antidecuplet masses can be expressed through
the physical ones and the mixing angles:
\bea
\n
N_8&=&N_1\cos^2\theta_N+N_2\sin^2\theta_N,\\
\n
N_{\at}&=&N_1\sin^2\theta_N+N_2\cos^2\theta_N,\\
\n
\Sigma_8&=&\Sigma_1\cos^2\theta_\Sigma
+\Sigma_2\sin^2\theta_\Sigma,\\
\la{NSbare}
\Sigma_{\at}&=&\Sigma_1\sin^2\theta_\Sigma
+\Sigma_2\cos^2\theta_\Sigma.
\eea
These masses must be substituted into \eq{GMO1} in order to
obtain the GMO relation for physical (mixed) members of the
octet and the antidecuplet:
\bea
\n
&&\frac{(N_1\cos^2\!\theta_N+N_2\sin^2\!\theta_N)+\Xi_8}{2}\\
\la{GMO2}
&=&
\frac{3\Lambda
+(\Sigma_1\cos^2\!\theta_\Sigma+\Sigma_2\sin^2\!\theta_\Sigma)}
{4}.
\eea
In addition one has relations for physical masses,
following from the originally equidistant spectrum of the
antidecuplet,
\bea
\n
z&=&N_1\sin^2\theta_N+N_2\cos^2\theta_N-\Theta\\
\n
&=&\!(\Sigma_1\sin^2\!\theta_\Sigma
+\Sigma_2\cos^2\!\theta_\Sigma)
-(N_1\sin^2\!\theta_N+N_2\cos^2\!\theta_N)\\
\la{GMO3}
&=&\Xi_{\frac{3}{2}}
-(\Sigma_1\sin^2\theta_\Sigma+\Sigma_2\cos^2\theta_\Sigma),
\eea
and a relation between $N,\Sigma$ mixing angles
\beq
\la{angles}
(N_2-N_1)\sin (2\theta_N)=(\Sigma_2-\Sigma_1)\sin(2\theta_\Sigma)
\quad(=-2V).
\eeq
[In the last equation one has to change the sign of the r.h.s.
if $(N_{\at}-N_8)(\Sigma_{\at}-\Sigma_8)<0$.]
A consequence of \eqs{GMO2}{GMO3} is a relation between physical
masses, independent of the mixing angles:
\beq
2(N_1+N_2+\Xi_8)=\Sigma_1+\Sigma_2+3\Lambda+\Theta.
\la{GMO4}\eeq

\Eqss{GMO2}{GMO4} generalize GMO relations for the octet
and the antidecuplet in case there is an arbitrary mixing
between them; these equations are analogous to \eq{GMOmes}
for strongly mixed mesons.

Recently Jaffe and Wilczek \cite{JW} suggested a
particular model of a strong antidecuplet-octet mixing with the
following mass formula:  $N_1\!=\!M_0,\, \Theta\!=\!M_0+m_s,\,
\Lambda=\Sigma_1 =M_0+m_s+\alpha,\, N_2\!=\!M_0+2m_s+\alpha,\,
\Xi_8\!=\!\Xi_{\frac{3}{2}}\!=\!M_0+2m_s+2\alpha,\,
\Sigma_2\!=\!M_0+3m_s+2\alpha.$ This model corresponds to a particular
choice of parameters in our general scheme:
$M_8\!=\!M_{\at}\!=\!M_0+\frac{1}{3}(5m_s+4\alpha),\,
x=\frac{1}{3}(2m_s+\alpha),\, y=\frac{1}{12}(-2m_s+5\alpha),\,
z=\frac{1}{3}(m_s+2\alpha),\, V=\frac{\sqrt{2}}{3}(2m_s+\alpha).$
It corresponds to the ``ideal mixing": $\theta_N=-\theta_\Sigma=
35.26^{\rm\small o}$.

The ideal mixing model, motivated by the hypothesis of
strong diquark correlations in the pentaquark baryons, is however not
too realistic.  First, it sets $\Lambda=\Sigma_1$, whereas the
values preferred by the PDG are $\Lambda=1600\,{\rm MeV}$ and
$\Sigma_1=1660\,{\rm MeV}$, although with significant error bars.
Second, it leads to $\Xi_8=1760\,{\rm MeV}$, whereas the only
candidate for the octet $\Xi$ known today is $\Xi_8=1690\,{\rm MeV}$.
Third, if the NA49 value of $\Xi_{\frac{3}{2}}=1862\,{\rm
MeV}$ is confirmed, it is very far from $\Xi_8$; Jaffe and
Wilczek have them degenerate.

Since the Jaffe--Wilczek model is a particular case of our general
formulae for mixing, we can try to get more realistic masses.
Assuming that $\Xi_{\frac{3}{2}}(1862)$ is the
member of the same antidecuplet as $\Theta^+$, we can use the
additional information, namely the position of $N_{\at}$ and
$\Sigma_{\at}$ \ur{NSzero} before the mixing is taken into account.
It means that actually the number of unknown variables ({\it i.e.}
the mixing angles $\theta_{N,\Sigma}$ and the masses of the
mainly-antidecuplet baryons $N_2$ and $\Sigma_2$) is the same as the
number of equations, so that they can be determined
from the masses of the mainly-octet baryons
$N_1\!=\!N(1440),\,\Lambda(1600),\,\Sigma_1\!=\!\Sigma(1660)$ and
$\Xi(1690)$.  Taken literally, however, these numbers do not lead to
any reasonable solution of \eqss{GMO2}{angles}. Solutions appear if
one shifts somewhat the above PDG-preferred values inside the error
bars.  For example, taking $N_1\!=\!1470,\, \Lambda\!=\!1570,\,
\Sigma_1\!=\!1635,\, \Xi\!=\!1700$ and solving \eqss{GMO2}{angles}
we obtain the following mixing angles and the masses of
mainly-antidecuplet baryons:
\bea
\n
\theta_N&=&13.1^{\rm\small o},\quad
\theta_\Sigma=19.1^{\rm\small o},\\
N_2&=&1656\,{\rm MeV},\quad\Sigma_2=1768\,{\rm MeV}.
\la{1sol}\eea
Another choice,
$N_1\!=\!1460,\, \Lambda\!=\!1575,\, \Sigma_1\!=\!1630,\,
\Xi\!=\!1710$, gives
\bea
\n
\theta_N&=&21.9^{\rm\small o},\quad
\theta_\Sigma=31.1^{\rm\small o},\\
N_2&=&1676\,{\rm MeV},\quad\Sigma_2=1799\,{\rm MeV}.
\la{2sol}\eea
We see thus that there is an uncertainty in predicting
the $N_2,\Sigma_2$ masses, which is due to the experimental
uncertainty in the mainly-octet masses. A general feature is that
mixing pushes both $N_2$ and $\Sigma_2$ masses to somewhat higher
values than those following from the antidecuplet equidistant
splitting, \eq{NSzero}. It is interesting to note that the recent
search for narrow resonances
in the partial-wave analysis of the elastic $\pi N$ scattering
\cite{Azimov} points out a candidate with the mass 1680 MeV.

We remind that we have ``pulled out" the
$N(1710)$ and $\Sigma(1880)$ resonances as presumably belonging to
another octet, therefore we are now discussing new unobserved
resonances.  The larger the mixing angle, more strangeness is
``washed out'' from the Roper resonance $N(1440)$ and brought into
the second nucleon $N_2$ which will then have a larger branching
into $N\eta$ decay. As the mixing angle rises, the mass of $N_2$
rises too, and the resonance becomes broader as it
mixes up with an extremely broad resonance $N(1440)$. Since no
broad $\half^+$ nucleon resonance is known in this mass range, it
means that mixing cannot be large and hence $N_2$ cannot be
considerably heavier than the `equidistant' value \ur{NSzero}. Our
``educated guess'' is that the new N resonance must be in the
1650-1690 MeV range and the new $\Sigma$ resonance must be in the
1760-1810 MeV range.

We note on this occasion that there is a weak (one-star) evidence for
a $\half^+$ $\Sigma$ resonance at 1770 MeV with a width of $70\,{\rm
MeV}$ in the PDG baryon listings: it might be a ``crypto-exotic''
partner of the more pronounced exotic $\Theta$ and $\Xi_{\frac{3}{2}}$.
Just because it is not exotic the $\Sigma$ resonance from the
antidecuplet can and probably does mix with $\Sigma$ from a nearby
octet and hence must be more broad than the narrow $\Theta$ and
$\Xi_{\frac{3}{2}}$. The same is true for the last, nucleon member of
the antidecuplet.

Is the value of the spacing in the antidecuplet $z=108\,{\rm MeV}$
reasonable? In ref. \cite{97} the spacing has been estimated as
$180\,{\rm MeV}$ using the evaluation \cite{GLS} of $\Sigma=45\,{\rm
MeV}$ for the nucleon sigma term. Since then $\Sigma$ has been
re-evaluated with a significantly larger result $\Sigma=67\pm 6\,{\rm
MeV}$ \cite{Arndt}. If one uses this value the antidecuplet splitting
reduces considerably. Let us remind a few simple equations from ref.
\cite{97}. The splittings in the ground-state octet, decuplet and
antidecuplet have been determined there by three constants
$\alpha,\beta,\gamma$ such that
\bea
\la{abg}
x&=&\frac{1}{10}\alpha+\frac{3}{20}\gamma,\qquad
y=\frac{1}{8}\alpha+\beta -\frac{5}{16}\gamma,\\
\n
z&=&\frac{1}{8}\alpha+\beta+\frac{1}{16}\gamma,
\qquad \alpha+\beta=\frac{2}{3}\frac{m_s}{m_u+m_d}\Sigma,
\eea
where $x,y$ characterize the splitting in the
octet and the decuplet (see \eqs{split810}{split_at}).
\Eq{abg} has been derived in the leading and subleading orders in
the number of colors $N_c$, implying $y\!=\!y^\prime$ which is well
satisfied. The combined least-square fit to the masses of the octet and
the decuplet gives $x=37.13,y=145.9,\sqrt{\sigma^2}=4\,{\rm MeV}$. We
take the quark mass ratio from the Gell-Mann--Oakes--Renner formulae
$m_s/(m_u+m_d)=12.9$ and the nucleon sigma term $\Sigma=72\,{\rm
MeV}$. We get then from \eq{abg} $z=108.8\,{\rm MeV}$ which, being
tripled, is compatible with the mass difference
$\Xi_{\frac{3}{2}}(1862)\!-\!\Theta(1539)$. It should be noted that
there can be certain $1/N_c^2$ corrections to \eq{abg}.
In short, one can
explain
the $\Xi_{\frac{3}{2}}\!-\!\Theta$ splitting.

Assuming $\Xi_{\frac{3}{2}}$ and $\Theta$ are members
of the same $\half^+$ antidecuplet, their widths are related by
$SU(3)$ Clebsch--Gordan coefficients (up to $O(m_s)$ corrections), and
proportional to the cube of meson momenta. Adjusting the equations
from ref. \cite{97} to the observed masses, we estimate
\bea
\n
\Gamma(\Xi_{\frac{3}{2}}^{--}\to\Sigma^- K^-)
&\approx& 0.87\,\Gamma_\Theta,\\
\Gamma(\Xi_{\frac{3}{2}}^{--}\to\Xi^- \pi^-)
&\approx& 1.68\,\Gamma_\Theta.
\la{gammaXi}\eea
Many-body decays of $\Xi_{\frac{3}{2}}$ add potentially at least
as much to the total width which we thus estimate as
$\Gamma_{\Xi_{3/2}}\simeq 5\Gamma_\Theta$. The restriction
$\Gamma_{\Xi_{3/2}}<18\,{\rm MeV}$ \cite{CERN} is consistent
with the most stringent experimental limit on the $\Theta$'s
width $\Gamma_\Theta<9\,{\rm MeV}$ \cite{ITEP} and with even more
restrictive indirect estimates of a few MeV or less \cite{narrow}.

Why are $\Theta^+$ and $\Xi_{\frac{3}{2}}$ so narrow? The small
width prediction of ref. \cite{97} is rather technical and calls
for a ``physical" explanation. To answer the question one has first
to explain why pentaquark baryons are unexpectedly light. Indeed,
a naive constituent quark model with quark mass
$M\approx 350\,{\rm MeV}$ (such that vector mesons are approximately
twice and nucleons thrice this mass) predicts $\Theta^+$ at
about $350\times 5=1750\,{\rm MeV}$, plus $100\!-\!150\,{\rm MeV}$
for strangeness.

The crucial physics ignored in this naive estimate is
the spontaneous breaking of chiral symmetry in QCD. It is due to
the SBCS that the nearly massless quarks of the QCD Lagrangian
obtain a large mass of about $350\,{\rm MeV}$. Simultaneously,
the massive (``constituent") quarks must interact
with light pions and kaons, and very strongly \cite{DP}. Chiral
forces are probably more important for binding constituent quarks
in baryons, than any other force~\cite{DPShif}.

In this picture, pentaquarks are baryons in which the additional
quark-antiquark pair appears in the form of an excitation of the
chiral field inside baryons. (It is not a pseudoscalar meson -
baryon molecule though.) The energy penalty for an additional pair is
not twice the constituent quark mass but something much less: it
is the energy of the chiral excitation, which is proportional to the
inverse size of the baryon. It tends to zero in the imaginary limit of
large-size baryons. This is a qualitative explanation why exotic
baryons are anomalously~light.

Even if the three `valence' quarks are not too relativistic,
the quark-antiquark pair inside the pion or kaon field is always
relativistic. The non-relativistic wave-function description of
pentaquarks makes no sense. ``Measuring" the quark position
with an accuracy higher than the pion Compton wave length of
$1\,{\rm fm}$ produces a new pion, {\it i.e.} a new quark-antiquark
pair. What makes sense in this situation is
describing baryons in the
infinite momentum frame. In that frame
there can be no production and annihilation of quarks, and the baryon
wave function falls into separate sectors of the Fock space: three
quarks, five quarks, etc. The difference between the ordinary nucleon
and $\Theta^+$ is that the nucleon has a three-quark component (but
necessarily has also a five-quark component) while $\Theta$'s Fock
space starts from the five-quark component.

One can now consider the decay amplitude of $\Theta$ or
$\Xi_{\frac{3}{2}}$ into an octet baryon and a pseudoscalar meson.
Owing to the Goldberger--Treiman relation it is equivalent to
evaluating the matrix element of the axial charge between the
antidecuplet and the octet baryons. In the infinite momentum frame
the axial charge does not create or annihilate quarks but only
measures the transition between the existing quarks. Therefore, the
matrix element in question is non-zero only between the pentaquark and
the {\em five-quark} component of the octet baryon. Hence it is
suppressed to the extent the five-quark component of a nucleon is less
than its three-quark component. In the case of the antidecuplet there
is an additional suppression owing to the specific flavor
structure of the nucleon's five-quark component where the antiquark
is in a flavor-singlet combination with one of the four
quarks \cite{DPP}.  This is why, qualitatively, the exotic $\Theta^+$
and $\Xi_{\frac{3}{2}}$ are narrow.

To summarize: the re-evaluation of the experimental nucleon sigma
term \cite{Arndt} leads to the re-evaluation of the splitting
inside the exotic antidecuplet of baryons \cite{97}, such that
the antidecuplet spacing of about $108\,{\rm MeV}$ as suggested by
the recent observation of the exotic $\Xi_{\frac{3}{2}}$ baryon
by the NA49 collaboration, can be accommodated. The
antidecuplet implies that there are crypto-exotic partners $N$
and $\Sigma$ of the exotic $\Theta^+$ and $\Xi_{\frac{3}{2}}$.
Since $N$ and $\Sigma$ have `usual' quantum numbers they can
in principle mix up with members of octets. One octet
in the mass range in question is nicely packed and satisfies
well the Gell-Mann--Okubo mass relation, therefore it can hardly
mix up strongly with anything else. The other octet [$N(1440),\,
\Lambda(1600),\,\Sigma(1660),\,\Xi(1690)$] violates the
Gell-Mann--Okubo relation and is a good candidate for mixing with
the antidecuplet, however the mixing cannot be too strong either.
Taking into account the possible mixing we arrive to the
conclusion that the $N$ and $\Sigma$ partners of the exotic
$\Theta$ and $\Xi_{\frac{3}{2}}$ are in the range 1650-1690 and
1760-1810 MeV, respectively. The higher they appear inside these
ranges, the broader these resonances must be. There is a weak
evidence of $\Sigma(1770)$ which could be a candidate for
the mainly-antidecuplet partner of exotic baryons but the nucleon
partner has to be searched for.

All numbers discussed in this paper are based on the assumption that
the mass of $\Xi_{\frac{3}{2}}$ is what has been found in the first
observation \cite{CERN} and that it will be confirmed by independent
experiments.

We thank Yakov Azimov and Maxim Polyakov for helpful discussions.
This work has been supported in part by the grant RFBR-0015-9606.

\end{document}